\documentstyle[12pt,epsfig]{article}
\newcommand{\STRUT}{\rule{0in}{2.0ex}}

\begin{document}
\vspace*{-0.6in}
\hfill \fbox{\parbox[t]{1.53in}{LA-UR-99-4309-REV}}
\vspace*{0.6in}

\begin{center}

{\Large {\bf Quartet n-d Scattering Lengths}}\\

\vspace*{0.20in}
by\\
\vspace*{0.20in}

J.\ L.\ Friar and D.\ H\"uber\\
Theoretical Division \\
Los Alamos National Laboratory \\
Los Alamos, NM  87545 \\

\vspace*{0.20in}
and\\
\vspace*{0.20in}

H.\ Wita\l a\\
Institute of Physics\\
Jagellonian University\\
PL-30059 Cracow, Poland\\

\vspace*{0.20in}
and\\
\vspace*{0.20in}

G.\ L.\ Payne\\
Department of Physics and Astronomy\\
University of Iowa\\
Iowa City, IA 52242\\

\end{center}
\vspace*{0.25in}

\begin{abstract}
Quartet n-d scattering lengths are calculated using second-generation
nucleon-nucleon potential models. These results are compared to the
corresponding quantity recently calculated using chiral perturbation theory.
\end{abstract}

\pagebreak

Solving exact few-body equations offers a possibility 
to test the present understanding of nuclear forces by direct 
comparison of theoretical predictions with experimental data. 
It is the scattering problem which provides the real opportunity to explore 
in depth the accuracy of our knowledge of the nucleon-nucleon 
interaction. Neutron-deuteron (n-d) elastic scattering at zero incident energy 
is the simplest three-nucleon scattering problem. At this energy 
only the s-wave scattering lengths survive. In the limit of 
relative n-d momentum $q_0 \to 0$ the eigenphase shift in the total 
angular momentum 3/2 state can be written in terms of quartet n-d 
scattering length $a_4$ by
\begin{eqnarray}
 \delta_4(q_0) \to -a_4\, q_0.
\label{eq1}
\end{eqnarray}
Accurate calculations of n-d quartet scattering lengths were first performed 10
years ago\cite{a4}. This quantity is known to be insensitive to most physics, 
such as $\ell > 0$ partial waves of the nucleon-nucleon (NN) potential and 
three-nucleon forces, because of constraints arising from the Pauli principle.  
The low (actually, zero) energy of the incoming neutron emphasizes s-waves, 
while the quartet spin emphasizes $S = 1$ between the two neutrons, which 
combination is Pauli forbidden. This reaction at zero energy depends only on 
details of the deuteron s-wave for an accurate calculation.

The potentials of a decade ago (sometimes called ``first-generation'' 
potentials) were not particularly accurate fits to the NN data base (or even to
the data bases in use when those potentials were constructed).  Deuteron
properties, such as binding energies and asymptotic normalization constants, had
considerable variations.  Thus, it is not surprising that three-nucleon
properties showed considerable spread due to these indifferent fits, although it
was never clear in advance which properties were suspect.  One such property was
$a_4$, the n-d quartet scattering length, where values of 6.304 fm and 6.380 fm
were obtained\cite{a4} for the RSC\cite{RSC} and AV14\cite{AV14} potential 
models, respectively.  Variations of these numbers due to partial-wave 
limitations or three-nucleon forces are of the order of $10^{-3} a_4$ (or 
less), which is much smaller than the potential-model difference. 
Such minimal influence of three-nucleon force effects and higher 
nucleon-nucleon partial waves is due to the fact that Pauli repulsion 
for three nucleons in the same spin state keeps the nucleons apart.

Recently, a new class of potentials has been developed (sometimes called
``second-generation'') that provides greatly improved fits to the NN data 
base\cite{2nd,AV18}. Only a single calculation\cite{pisa} of $a_4$ exists for 
a single second-generation potential model (AV18)\cite{AV18}, and that result 
lies between the RSC and AV14 results listed above.  Until very recently, no 
particular motivation existed for revisiting the $a_4$ calculations.

Chiral perturbation theory\cite{cptnuc} ($\chi$PT) provides an alternative path
(to conventional potentials) for calculating few-nucleon observables. Scattering
amplitudes are constructed directly from a field theory, employing one or 
another scheme of regularization and renormalization. In this fashion the first
three-nucleon calculation exploiting chiral perturbation theory was recently 
performed\cite{cpta4} for the observable $a_4$.  The result, 6.33(10) fm, lies
between the RSC and AV14 results quoted above, which motivates this brief update
of the theoretical situation.

\begin{table}[htb]
\centering
\caption{Quartet $n$-$d$ scattering lengths ($a_4$, in fm)  calculated using
potential models and $\chi$PT, together with the experimental value.}

\hspace{0.25in}

\begin{tabular}{|l||ccccc||c||c|}
\hline
type\rule{0in}{2.5ex} & N93\cite{2nd}& N II\cite{NII} & RSC93\cite{2nd} &
CDB\cite{CDB} & AV18\cite{AV18}&$\chi$PT\cite{cpta4}&Expt.\cite{expt}\\ 
\hline \hline
$a_4\STRUT$&6.346&6.343&6.353&6.350&6.339&6.33(10)&6.35(2)\\ 
\hline
\end{tabular}
\end{table}

We have calculated $a_4$ for a variety of second-generation NN potentials listed
in Table I.  These include the Nijmegen 93 (N93; nonlocal), the Nijmegen II 
(N II; partial-wave local), the Reid soft core 93 (RSC93; partial-wave local),
the CD-Bonn (CDB; nonlocal), and the Argonne V18 (AV18; local) potentials.  The
large difference ($>$1\%) seen between the previous (first-generation)
potential-model results is not reproduced in our five (second-generation)
results, which are within a factor of $2 \cdot 10^{-3}$ of each other. We also
note the the AV18 potential contains an electromagnetic force that must be
turned off in momentum-space procedures in order to obtain a result. We have
determined using a configuration-space approach\cite{a4} that eliminating this
force component lowers $a_4$ by approximately 0.018 fm, which is a very small
change. Our result in Table~1 incorporates the complete force, and is slightly
larger than that of Ref.~\cite{pisa}. All (second-generation) theoretical 
results agree with the experimental value.

The large discrepancy seen for first-generation potentials has vanished. 
Second-generation potential results are now in close agreement with the 
$\chi$PT result.  Although the latter has a relatively large theoretical error 
bar, that error reflects an estimate of uncalculated higher-order Lagrangian
terms.  Given that these would roughly correspond to small components of the
nuclear potential (which scarcely affect the result), it seems likely that the
error is overestimated for this reaction.

In summary, second-generation NN potential calculations of $a_4$ are in much
better agreement with each other, and with chiral perturbation theory, than were
older first-generation potential calculations.

\pagebreak

\begin{center}
{\large {\bf Acknowledgments}}\\
\end{center}

The work of JLF and DH was performed under the auspices of the United States 
Department of Energy, while that of GLP was supported in part by the United 
States Department of Energy. The work of DH was supported in part by the
Deutsche Forschungsgemeinschaft under Project No. Hu 746/1-2. The work of HW was
supported by the Polish Committee for Scientific Research.  One of us (JLF)
would like to thank G.\ M.\ Hale of Los Alamos for helpful discussions about n-d
scattering.

\end{document}